\documentclass[aps,twocolumn,showpacs,amsmath,amssymb,superscriptaddress,10pt,floatfix]{revtex4-1}

\usepackage{amsmath}
\usepackage{amsfonts}
\usepackage{amssymb}
\usepackage{color}
\usepackage{verbatim}
\usepackage{graphicx}
\usepackage{soul}

\newcommand{\beq}{\begin{equation}}
\newcommand{\eeq}{\end{equation}}
\newcommand{\bqa}{\begin{eqnarray}}
\newcommand{\eqa}{\end{eqnarray}}

\newcommand{\erf}[1]{Eq.~(\ref{#1})}

\newcommand{\bra}[1]{\left\langle{#1}\right|}
\newcommand{\ket}[1]{\left|{#1}\right\rangle}

\newcommand{\sq}[1]{\left[ {#1} \right]}

\newcommand{\blk}{\color{black}}

\definecolor{ngreen}{rgb}{0.2,0.6,0.2}

\definecolor{purp}{rgb}{0.8,0.1,0.8}

\definecolor{golden}{rgb}{0.8,0.6,0.1}

\newcommand{\LX}[0]{\overleftarrow{X}}
\newcommand{\RX}[0]{\overrightarrow{X}}
\newcommand{\lx}[0]{\overleftarrow{x}}

\newcommand{\cc}{C _ {\rm c}}

\renewcommand{\section}[1]{\textit{#1}.---}

\begin{document}

\title{Experimental quantum processing enhancement in modelling stochastic processes}

\author{Matthew S. Palsson}
\affiliation{Centre for Quantum Computation and Communication Technology (Australian Research Council), Centre for Quantum Dynamics, Griffith University, Brisbane, 4111, Australia}

\author{Mile Gu}
\affiliation{School of Physical and Mathematical Sciences, Nanyang Technological University, Singapore 639673, Republic of Singapore}
\affiliation{Complexity Institute, Nanyang Technological University, 60 Nanyang View, Singapore 639673, Republic of Singapore}
\affiliation{Centre for Quantum Technologies, National University of Singapore, 3 Science Drive 2, Singapore, Republic of Singapore}

\author{Joseph Ho}
\affiliation{Centre for Quantum Computation and Communication Technology (Australian Research Council), Centre for Quantum Dynamics, Griffith University, Brisbane, 4111, Australia}

\author{Howard M. Wiseman}\email{H.Wiseman@griffith.edu.au}
\affiliation{Centre for Quantum Computation and Communication Technology (Australian Research Council), Centre for Quantum Dynamics, Griffith University, Brisbane, 4111, Australia}

\author{Geoff J. Pryde}\email{G.Pryde@griffith.edu.au}
\affiliation{Centre for Quantum Computation and Communication Technology (Australian Research Council), Centre for Quantum Dynamics, Griffith University, Brisbane, 4111, Australia}

 \maketitle

\noindent \textbf{Computer simulation of observable phenomena is an indispensable tool for engineering new technology, understanding the natural world, and studying human society~\cite{background}. Yet the most interesting systems are often complex, such that simulating their future behaviour demands storing immense amounts of information regarding how they have behaved in the past. For increasingly complex systems, simulation becomes increasingly difficult and is ultimately constrained by resources such as computer memory. Recent theoretical work shows quantum theory can reduce this memory requirement beyond ultimate classical limits~\cite{mgu} (as measured by a process' statistical complexity~\cite{crutch09}, $C$). Here we experimentally demonstrate this quantum advantage in simulating stochastic processes. Our quantum implementation observes a memory requirement of $C_q = 0.05 \pm 0.01$, far below the ultimate classical limit of $C = 1$. Scaling up this technique would substantially reduce the memory required in simulation of more complex systems.}

What new tasks can be enhanced by quantum information science?  It is a matter of practical importance and  fundamental interest to find new additions to the impressive list of known quantum information benefits that include: the exponential speed-up provided by Shor's factorisation algorithm \cite{shor} and by algorithms for simulating quantum systems \cite{dominic}; the physically guaranteed security of quantum key distribution \cite{qkd}; and the sensitivity advantages in using certain quantum states for metrology \cite{metrology1,metrologyrev}. In this work we experimentally demonstrate a fundamentally new quantum advantage: quantum information processing can reduce the memory required to simulate a stochastically evolving classical system, by encoding information in nonorthogonal quantum states. Limitations on memory availability are a key consideration in computer simulation, as the state space grows exponentially with the size of the system.

Our work is of particular relevance to the field of complexity theory. There,
the phenomena that people seek to understand --- such as neural networks or the dynamics of the stock market ---  consist of a vast myriad of interacting components, whose internal details are too complex or inaccessible for one to model their behaviour from first principles. In such cases, the system is instead typically regarded as black box, such that one has access only to some observable output. The task is then to isolate key indicators of future behaviour from this data --- and this data alone --- without any knowledge of the system's internal mechanism. It is possible to imagine that many different models of this type could be constructed for a given process. Of these, simpler models --- those that store less data without sacrificing predictive accuracy --- then represent a better understanding of exactly what observations in the past matter for the future. Our experimental work aims to demonstrate that, in taking this motivation to its ultimate conclusion, quantum effects can provide a powerful resource for simplifying models. 

Our technique is fundamentally different from quantum data compression~\cite{Plesch2010,Rozema2014}. The former is concerned with preserving all input data, and thus encodes orthogonal signal states into orthogonal encoded states. By contrast, our work is concerned with more efficient ways of discarding useless data (in the sense of being useless for future prediction), by encoding classically distinct states as non-orthogonal quantum states, and processing them coherently.

 To demonstrate the quantum advantage provided for this kind of simulation task, we need to quantify the minimum amount of memory --- i.e.\ stored information --- required to simulate a process. Mathematically, we can characterize the observable behavior of a dynamical process by a joint probability distribution $P(\LX,\RX)$, where $\LX$ and $\RX$ respectively represent random variables that govern the observed behavior of the process in the past and future.  A \textit{simulator}, implementing a model for the process, operates by storing information about $\LX$ within some physical system $\mathcal{S}$, such that for each instance of the process with a particular past $\lx$, it can be set to a particular state that allows reproduction of expected future statistics, i.e.\ generate a random variable sampled from $P(\RX |\LX = \lx)$.

The complexity of the simplest simulator - the one for which $\mathcal{S}$ has minimal entropy - is regarded as an intrinsic property of the process being simulated, capturing the bare minimum information one must store to replicate the statistics of the process \cite{shal01,crutch09}.  In complexity theory, the minimal entropy of $\mathcal{S}$ is known as the statistical complexity $C$. The most complex processes reside between complete randomness (maximum system entropy) and complete order (zero system entropy)~\cite{hog}.  At each extremity, the entropy of the \textit{simulator} is zero: $C=0$.  The statistical complexity  has  been applied to  a wide range of problems, including self organisation \cite{Shal04}, the onset of chaos \cite{crutch89} and the complexity of a protein configuration space \cite{chun08}.

The statistical complexity of a stochastic process can be determined by dividing the set of all possible pasts into equivalences classes, such that all members of a given class yield coinciding future predictions. The implementation of such a model can replicate future statistics by recording only which equivalence class $s$ that $\lx$ belongs to. In the literature, these equivalence classes are known as causal states \cite{crutch89}; thus, causal states encode the information that is required to be stored. The complexity of such a simulator is then given by its entropy \cite{Schumacher}
\begin{equation} \label{Cc}
\cc = - \sum \wp_s \log \wp_s,
\end{equation}
where the sum is taken over all causal states $s \in S$, and $\wp_s$ is the probability that $\lx$ lies in $s$. This representation turns out be classically optimal \cite{crutch89}  ---  no classical model can simulate a stochastic process storing less memory than $C_{\rm c}$. Thus $C_{\rm c}$ coincides with the statistical complexity.

Na\"{i}vely, one might expect such optimal models to waste no information --- any information they store should be of relevance to the future. Surprisingly, this is not so. Classical models are almost always inefficient. Even in very simple processes, the statistical complexity $C_{\rm c}$ is generally strictly greater than $E = I(\LX,\RX)$, the mutual information between past and future outputs~\cite{crutch09}. Some information stored within a simulator is simply wasted. This surprising wastefulness of even the provably most efficient classical models can be very significant for more complex systems; and contributes to an unavoidable energy cost in stochastic simulation~\cite{karoline,andrew}.

Quantum information processing can drastically reduce this waste. It has been theoretically demonstrated that. for any process whose optimal classical model with $C_{\rm c} > E$, there exists a quantum model that requires a smaller memory, $C_{\rm q} < C_{\rm c}$~\cite{mgu}. Quantum models assign each causal state $s$ an associated quantum state $\ket{\tilde s}$. The quantum states in this set are, in general, mutually non-orthogonal, but nevertheless can be used to replicate desired future statistical behaviour. This nonorthogonality ensures that the size of the quantum memory required to retain $\ket{\tilde s}$---the von Neumann entropy of the mixed state $\rho = \sum_s \wp_s \ket{\tilde{s}}\bra{\tilde{s}}$ (ref.~\cite{Schumacher})---is \blk lower than its classical counterpart.

\begin{figure*}[h!]
 \centering
 \includegraphics[width=\textwidth]{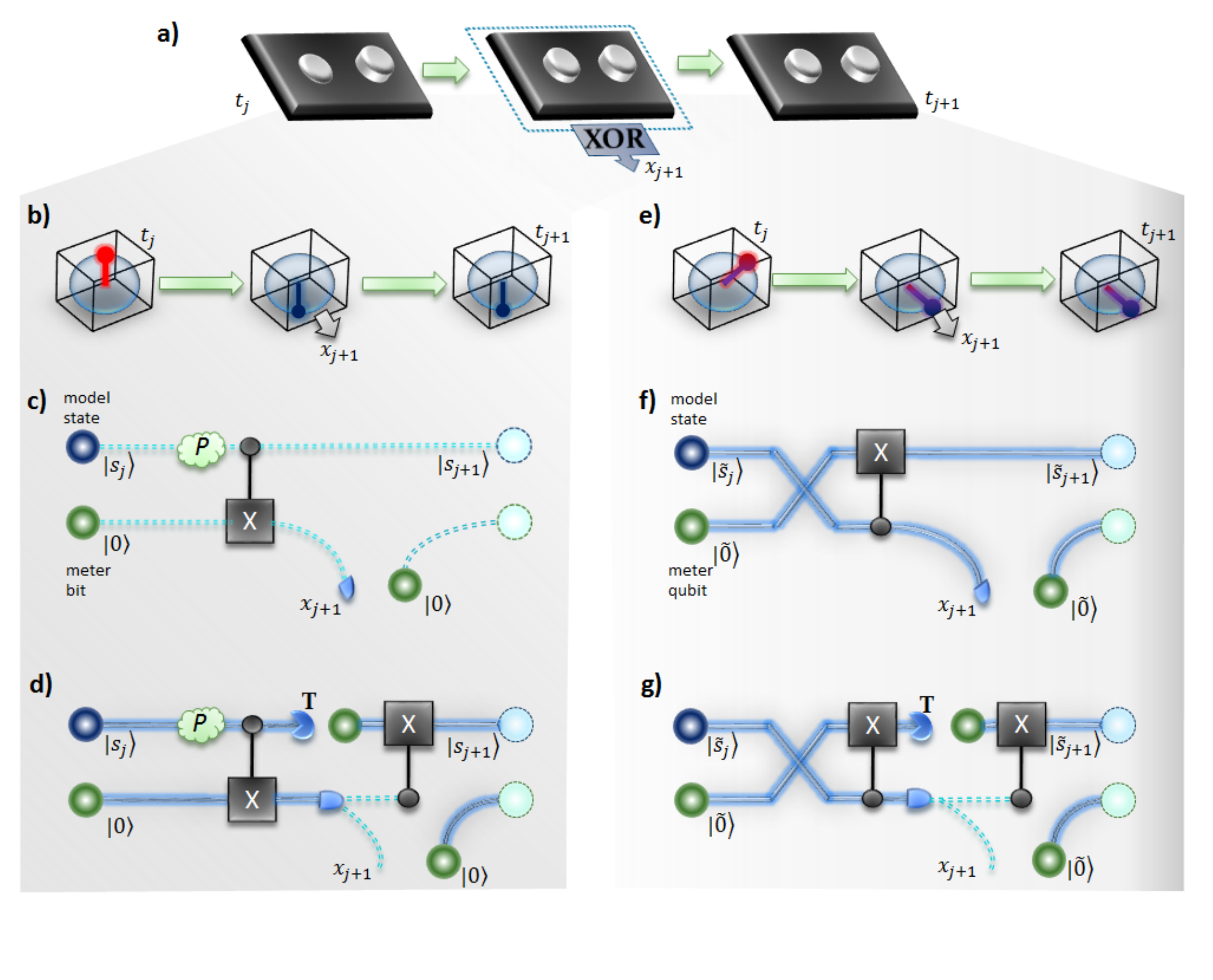}
 \caption{
 \textbf{Representation of a stochastic system, with classical and quantum statistical models, at the $(j+1)^{\rm th}$ time step of evolution.}
 \textbf{a.} The example system is a pair of switches whose settings determine the value of an output bit, and whose settings are randomized by a probabilistic process during the step (see text for details). \textbf{b.} Since the output is determined solely by the parity of the switches, a one-bit classical model can be used to represent the system and produce equivalent output statistics. In the example shown, the orientation (up or down) of the vector --- or equivalently its colour (red or blue) --- represents the state of the model and determines the output bit $x_{j+1}$. \textbf{e.} A quantum model allows for reduced complexity (see text for details) by encoding the state into nonorthogonal quantum states (the multi-colour vectors, with non-polar orientations, represent quantum superpositions of logical states). \textbf{c.} A conceptual classical circuit (double lines represent classical bit rails) for realizing the operation of the classical model above. The classical input state (top rail) at time $t_j$ is subjected to is a probabilistic action, potentially flipping the state. The model state is then correlated with a meter bit (bottom rail), initially in the logical zero state, via a CNOT gate. Reading out the meter via a logical (Pauli ``Z'') measurement provides the output bit. After readout, the model state is passed on to the next time step, along with a fresh meter bit.
 \textbf{f.} A conceptual quantum circuit (glowing lines represent qubit states) for realizing the quantum model above. The operation is similar to the classical circuit in (\textbf{c}), except that the probabilistic action is delayed until the read-out of the meter (as above), which yields a random result, and collapses the model state because of the entanglement
generated by the CNOT gate acting on superposition states. \textbf{e.} and \textbf{g.} Conceptual circuits of the classical and quantum models, as experimentally realized. The key difference (for practical reasons only) is the interruption of the model states for characterizing measurements (denoted T, for quantum state \textit{tomography}) with subsequent repreparation.
}
 \label{fig:models}
\end{figure*}

We experimentally demonstrate these ideas by modeling a specific, simple, stochastic process. It applies to many different physical systems, one of which is illustrated in Fig.~\ref{fig:models}\textbf{a}: a pair of binary switches. 
At each time step $j$, one of the switches is chosen at random and flipped with probability $P$. The system then outputs $0$ if the switches are aligned and $1$ if they are anti-aligned. The obvious (perhaps na\"{i}ve) model keeps track of the state of both switches, resulting in a memory of entropy~$2$. However, we may optimise this classical model by observing that the parity of the switches corresponds to the causal states of the system (any past histories for coinciding switch parity have statistically identical futures). Thus to simulate its statistics, we need only store a single binary value, $s$, that takes on $0$ and $1$ with equiprobability.

\begin{figure}
 \centering
\includegraphics[width=\columnwidth]{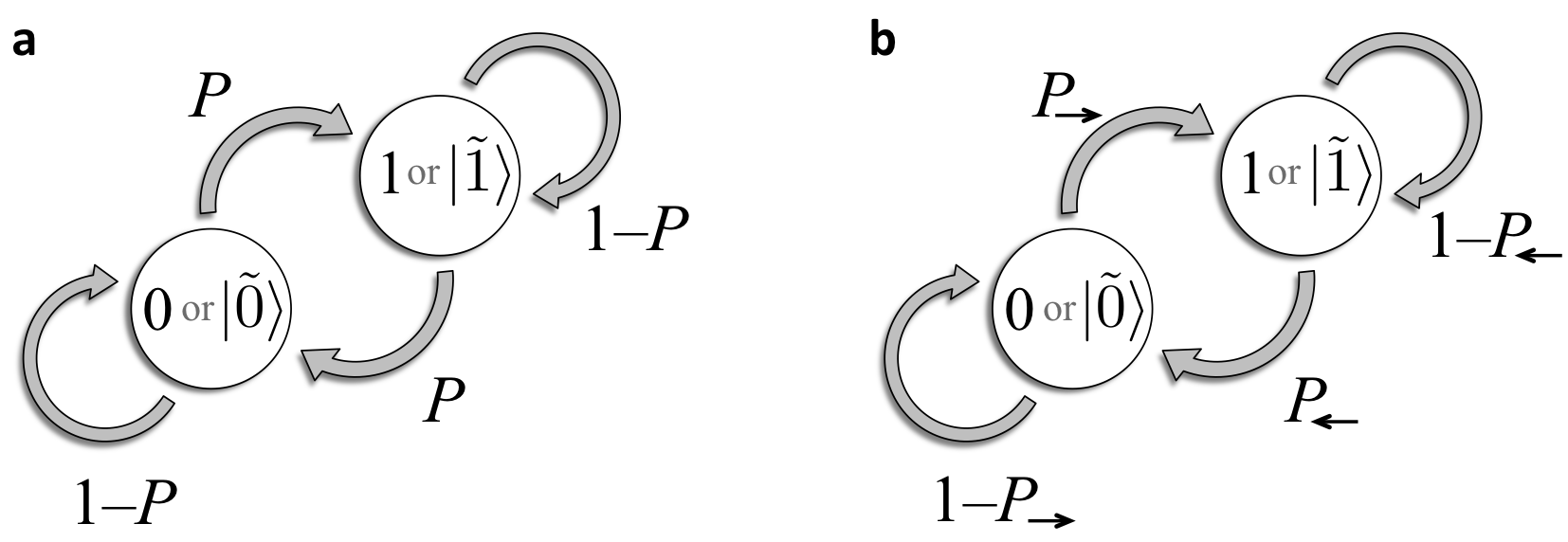}
 \caption{  \label{fig:transiton}\textbf{Replicating statistical behaviour with causal states --- transition diagram for the model.} \textbf{a.} In the example of Fig.~\ref{fig:models}, the probability of the model state bit transitioning from one causal state (denoted here by a circle) to the other is $P$, and thus the probability of remaining in the same state is $1-P$. \textbf{b.} In general, a two-causal-state model may have a transition probability, either $P_\rightarrow$ or $P_\leftarrow$, that depends on the causal state at the beginning of the step. The case we consider for most of this work, $P=P_\rightarrow=P_\leftarrow$, is a particular example.}

\end{figure}

Figure~\ref{fig:transiton}\textbf{a} summarises how the dynamics of this process is completely captured by transitions between the two causal states. The steady-state occupation probabilities $\wp_0$ and $\wp_1$ for the two causal states coincide due to symmetry. Thus this process, in general, has a statistical complexity of $\cc = 1$. The only exception is when when $P = 0.5$, where $\cc = 0$ since then each output bit is completely random (uncorrelated with earlier output bits), so that no memory is required. A potential representation of the simplest model is illustrated in Fig.~\ref{fig:models}\textbf{b}, which uses antiparallel red and blue \blk vectors to represent the two causal states.

Figure~\ref{fig:models}\textbf{e} provides a conceptual representation of the quantum causal states. The quantum model makes use of a non-orthogonal encoding, such that each of the two values of $s$ is assigned a quantum state $\ket{\tilde{s}}$, namely:
\begin{align}
 \ket{\tilde{0}} &= \sqrt{1-P}\ket{0}+\sqrt{P}\ket{1}  \label{eqn:tilde0}; \\
 \ket{\tilde{1}} &= \sqrt{P}\ket{0}+\sqrt{1-P}\ket{1}    \label{eqn:tilde1}.
 \end{align}
Here, $\ket{0}$ and $\ket{1}$ are the logical basis states of a qubit. The quantum-enhanced model saves further memory by sacrificing absolute knowledge of switch parity - it distinguishes the two possible immediate pasts only to the extent required to generate correct future statistics. The storage of $\ket{\tilde{s}}$ in a physical system $\mathcal{S}$, rather than the classical states $s$, results in an reduced simulator entropy of
\begin{align}
  C_{\rm q}  = -\text{Tr}(\rho \log_2 \rho),
\end{align}
where 
\beq \label{rho}
\rho = \frac{1}{2}\left(\ket{\tilde{0}}\bra{\tilde{0}}+\ket{\tilde{1}}\bra{\tilde{1}}\right)
 = \frac{1}{2}\sq{\hat 1 + 2\sqrt{P(1-P)}\hat X} \blk
\eeq
represents the state of $\mathcal{S}$ averaged over possible causal states, and $\hat X$ is the Pauli operator.
The coherence in \erf{rho} comes from
the nonorthogonality, which guarantees reduced complexity $C_{\rm q}<C_{\rm c}$ for any $P$ (except $P=0.5$, where $C_{\rm q}=C_{\rm c}=0$). For $P$ close to $0.5$, $C_{\rm q}$ can be arbitrarily small, while $C_{\rm c}=1$. This theoretically predicted behavior is plotted in Fig.~\ref{fig:UnBdata}.

Figures~\ref{fig:models}\textbf{c} and \ref{fig:models}\textbf{f} show, respectively, classical and quantum logical circuits that inplement these models.
The operation of the circuits is explained in detail in the caption, but the key point is that the $(j+1)^{\rm th}$ simulation step (going from discrete time $t_j$ to $t_{j+1}$, say) involves taking the memory as input, applying the probabilistic operation, and generating a classical output $x_{j+1}$. In the classical case the probability $P$ of a flip is inserted externally, but in the quantum case it comes from the intrinsic randomness
 of quantum measurements on nonorthogonal states.
In either scenario, the resulting predictive model can faithfully replicate future statistical behaviour. That is, when initialised in the appropriate (quantum) causal states at time $t$, the future outputs are statistically indistinguishable, and align with that of the original process being modelled.

\section{Experimental Implementation}
\begin{figure}
 \centering
\includegraphics[width=\columnwidth]{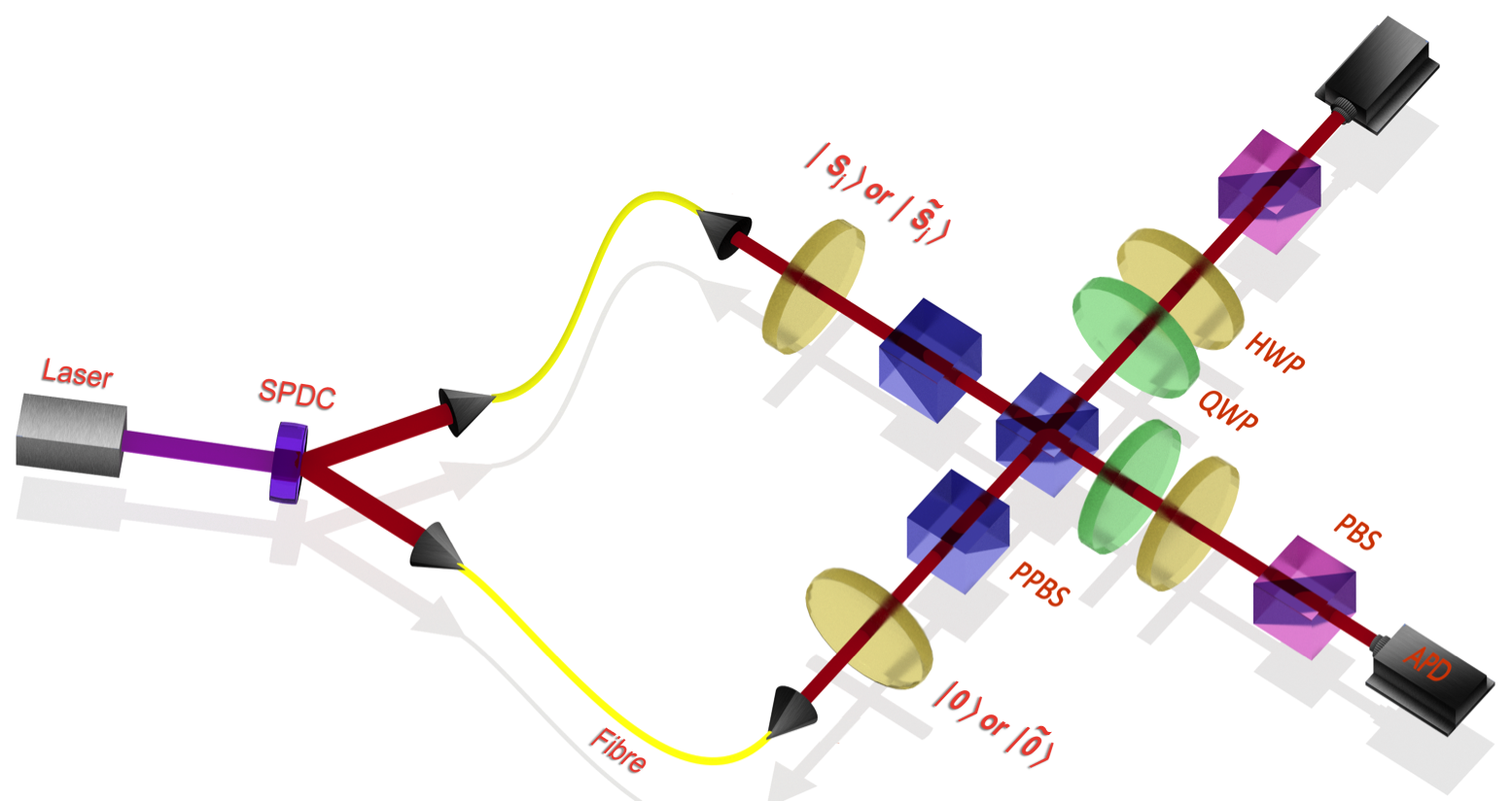}
 \caption{ \textbf{Experimental setup.} Photons from a cw-pumped spontaneous parametric downconversion (SPDC) source are prepared in the relevant input states by half wave plates (HWPs) and are incident on a linear optics controlled-NOT gate realized with partially polarizing beam splitters (see Methods). Converting between classical and quantum models requires changing the input states from classical (orthogonal) polarization states to nonorthogonal superposition states.
Measurement of one output determines the output bit at the current time step, the other output is tomographically characterized over many measurement runs to determine the state of the model and its entropy.
 Key elements include: polarising beam splitters (PBS), partially polarising beam splitters (PPBS), quarter and half wave plates (QWP \& HWP), and avalanche  photodiode (APD) single photon detectors.}
 \label{fig:gate}
\end{figure}

We implemented the quantum switch model using a photonic quantum logic circuit. We compared it with the theoretical classical bound and and a classical switch model that we also implemented with a photonic ciruit. 
Figs~\ref{fig:models}\textbf{d} and \ref{fig:models}\textbf{g} show the mapping of the conceptual models onto what we realised experimentally.
Experimentally processing either classical states (classical model) or quantum states (quantum model) required a controlled-NOT gate, as well as two single photons---one to encode the state of the model, and one to facilitate readout. We used a linear optics controlled-Z gate (Fig.~\ref{fig:gate}) with local unitary operations, and spontaneous parametric downconversion for photon generation, to realise these (see Methods).

 In the classical circuit, the causal states are encoded in orthogonal logical photon polarisation states, the equivalent of classical bits. The controlled-NOT gate performs a classical XOR operation, mapping the system state (after the probabilistic operation) onto a meter bit, which is read out via a
 destructive
 projective measurement to provide the $(j+1)^{\rm th}$ data value of the model output.

In the quantum circuit, the relevant quantum causal states are encoded in non-orthogonal photon polarisation states, as per Eqs~(\ref{eqn:tilde0},\ref{eqn:tilde1}). The controlled-NOT gate produces an entangled state between the model state and a meter qubit. The probability of a flip is
 determined by the degree of orthogonality of the causal states. Destructive projective measurement of one qubit after the CNOT gate produces a classical output which is the $(j+1)^{\rm th}$ data value of the model output, and the corresponding collapse of the quantum state on the other photon implements the probabilistic operation on the model qubit for the next time step.

In order to verify the statistical complexity of the simulation, we measure the entropy of the model register via quantum state tomography at the end of time-step circuit. This requires a destructive measurement of the photonic memory (qu)bit, and consequently repreparation using classical logic. This is a slight practical difference from the theoretical circuits of Figs~\ref{fig:models}\textbf{c} \& \textbf{f}, and is necessary only for the sake of verifying the information storage requirements of the quantum model.

\begin{figure}[h!]
 \centering
 \includegraphics[width=\columnwidth]{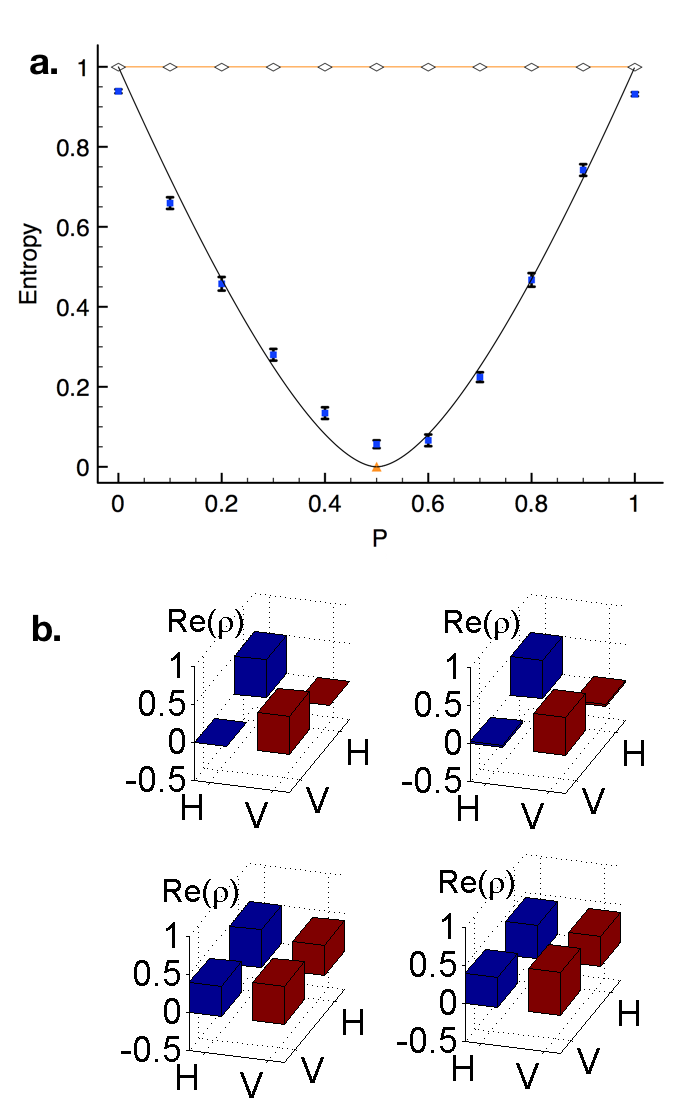}
 \caption{ \textbf{Experimental data for classical and quantum models of the stochastic process.} \textbf{a.} Experimentally measured statistical complexities (entropy) for the classical and quantum model states, sampled for  a range of values for $P\  (=P_\rightarrow=P_\leftarrow)$. Blue squares are the quantum data, black diamonds are the classical data. The orange solid line represents the theoretically calculated
 entropy for the classical  scenario.  The  black curve represents the theoretically-calculated entropy for the quantum  scenario. Error bars are one standard deviation, derived from Poissonian counting statistics. Error bars not shown are much smaller than the data points. The orange triangle denotes the classical prediction for $P=0.5$, where no memory is required for the corresponding completely random process.  \textbf{b.} Real parts of the tomographically determined equilibrium density matrix, at the model state output, for $P=0.8$. Top: classical model. Bottom: quantum model. Left: theory. Right: experiment. Imaginary components (small) are not shown.}
 \label{fig:UnBdata}
\end{figure}

Experimental determinations of the statistical complexity, for both classical and quantum models are shown in Fig.~\ref{fig:UnBdata}\textbf{a}. We collected  data for various values of $P$ ranging from $0$ to $1$ at intervals of $0.1$. For a wide range of $P$ values, $C_q^\textrm{exp} < C_c=1$, as predicted by theory. Small imperfections in $C_q^\textrm{exp}$ arise due to slight imperfections in the operation of the CNOT gate and preparation of the input states.
Figure \ref{fig:UnBdata}\textbf{b} shows theoretical and experimental single qubit density matrices, for the symmetric case where $P=0.8$, and provide a good example of the strong agreement between theory and experiment.

The classical scenario for this model uses orthogonal logical states, which are invariant with $P$.
Experimentally, this leads to a single data point for the classical symmetric case, since the statistics are independent of $P$.  Experimental imperfections, as discussed, led to a measured value of $C_c^\textrm{exp} = 0.9992\pm 0.0002$,  very slightly less than the predicted value of unity (Fig.~\ref{fig:UnBdata}\textbf{a}). The imperfections, at the $\lesssim 0.1\%$ level, bias the equilibrium statistics. Note that measuring a value less unity does not imply that the classical bound of unity is incorrect, but rather that our slightly imperfect experiment implements a classical model of a slightly different process, one with a statistical complexity marginally less than $1$.
 
Our setup can also be generalised to model a class of more general stochastic processes, including the case where probabilities of transitioning between the two causal states do not coincide (See Fig.~\ref{fig:transiton}\textbf{b}). This is the case, for example, when the probability of flipping a switch depends on its current parity. While the causal states of such a process remains unchanged, this generalisation does affect the transition probabilities between the two causal states - and thus their equilibirum distribution. In general, $\wp_1 \neq \wp_2$ and thus $C_{c} \neq 1$ (see Methods for details).

We experimentally tested one such case, where ${P_\leftarrow}=0.3$ and ${P_\rightarrow}=0.9$. The experimental implementation is the same as before, except that the states $\ket{\tilde{0}}$ and $\ket{\tilde{1}}$ are no longer symmetrically distributed about $\left(\ket{0}+\ket{1}\right)/\sqrt{2}$.
The experimentally determined entropy for the quantum model is $S_q=0.19\pm0.01$, much lower than the equivalent case for the classical model, $S_c=0.818\pm0.001$. Note that these values are slightly in excess of the theoretically predicted values of $0.12$ and $0.81$ respectively, which we attribute to a combination of slightly imperfect state preparation and slightly imperfect CNOT gate operation.

\section{Conclusions} In complexity theory, the statistical complexity of a stochastic process is considered as a intrinsic quantifier of its structure - representing the ultimate limit in the amount of memory required needed to optimally simulate its future statistics. Here, we have experimentally demonstrated that this limit can be surpassed with quantum processing. Stochastic processes permeate quantitative science, modelling diverse phenomena from neural networks to financial markets. In complexity theory, the construction of the simplest such models that replicate their observation behaviour has played an important role in understanding their hidden structure. Our results present a proof-of-principle that these existing methods can be enhanced through quantum technology. Recent theoretical work indicate quantum models can be further improved for non-markovian processes~\cite{Mahoney2015}, and our technology could be adapted to realising these designs.

As the amount of information classical models waste often scales with the complexity of the processes they model, the adoption of our methods could have significant potential in simplifying more complex simulations. This highlights quantum theory's relevance not only in understanding the microscopic world, but also its importance in studying the complex, macroscopic systems that are characteristic of everyday life.

\section{Methods}
\textbf{Photon source and CNOT gate.} A  source of  polarisation-unentangled  photon pairs was realised using type-I spontaneous parametric down conversion in Bismuth Borate (BiBO). The source produced photon pairs at 820~nm when pumped with a 410~nm, continuous-wave,  60~mW diode laser. The classical or quantum logic
 was  implemented  by constructing a linear optics CNOT gate as shown in Fig.~\ref{fig:gate}. 
 (In practice, the CNOT gate is realised using a controlled-Z (CZ) gate~\cite{langford} and Hadamard rotations which are incorporated into the settings of the wave plates before and after the gate.)  To determine how well the CNOT gate is operating~\cite{white}, we attempted to generate a maximally entangled state from separable inputs, with the resultant two-qubit density matrix reconstructed  via quantum state tomography 
  The fidelity of the state produced by the CNOT gate with the desired maximally entangled state was measured to be 0.97$\pm$0.01.

\textbf{Classical XOR gate.} The classical XOR gate, used to implement the model with classical causal states, can be implemented using a quantum CNOT gate (as above) and orthogonal logical photon polarisation states as bits. Due to experimental contingencies, we collected the classical data using a different (but nominally identical in layout and component type) CNOT gate to the one used for the quantum data collection, and at a later time.

\textbf{The asymmetric two-switch process.} In the main text, we studied the special case of the two-switch process where the probability of flipping a switch did not depend on the state of the two-switches. We can generalise this model by assuming that a switch is flipped with probability $P_\rightarrow$ when the switches align, or $P_\leftarrow$ otherwise (Fig.~\ref{fig:transiton}\textbf{b}). The resulting system will still have two causal states, $s = \{0,1\}$, corresponding to the parity of the two switches. The two causal states, however, no longer occur with equiprobability, and instead satsify $\wp_0 P_\rightarrow =\wp_1 P_\leftarrow$. Thus $C_c \leqslant 1$, with equality when $P_\rightarrow = P_\leftarrow$. The quantum model causal states for the general model is given by
\begin{align}
 \ket{\tilde{0}} &= \sqrt{1-P_\rightarrow}\ket{0}+\sqrt{P_\rightarrow}\ket{1}  \label{eqn:2},\\
 \ket{\tilde{1}} &= \sqrt{P_\leftarrow}\ket{0}+\sqrt{1-P_\leftarrow}\ket{1}    \label{eqn:3}.
 \end{align}
The statistical complexity is given by the von Neumann entropy of $\rho=\wp_0\ket{\tilde{0}}\bra{\tilde{0}} +
\wp_1\ket{\tilde{1}}\bra{\tilde{1}}$. The quantum circuit to realise this quantum model is the same as that in Fig.~\ref{fig:models}\textbf{f}, except that we must replace the controlled-$\hat{X}$ (CNOT) operation with a  controlled-$\hat{U}$ (CU) gate, where $\hat{U}$ is the operator such that $\hat{U}  \ket{\tilde{0}}  =  \ket{\tilde{1}}$. In practice, $\hat{U}=\hat{V}\hat{X}\hat{V}^\dag$, where $\hat{V}\left( P_\rightarrow,P_\leftarrow \right)$ is a rotation about the $Y$-axis of the Bloch sphere. Thus, the rotation can be implemented by HWPs in the meter arm before and after the CNOT gate --- we incorporate these rotations into the settings of the state preparation and measurement waveplates. However, the asymmetry of settings, when implemented with experimental components, leads to a slight degradation in the performance of the CU gate compared with the CNOT case.

\section{Acknowledgements}
This research was funded in part by the Australian Research Council Centre of Excellence for Quantum Computation and Communication Technology (Project number CE110001027). MG is financially supported by the John Templeton Foundation Grant 53914 {\em ``Occam's Quantum Mechanical Razor: Can Quantum theory admit the Simplest Understanding of Reality?''} and  National Research Foundation (NRF), NRF-Fellowship (Reference No: NRF-NRFF2016-02).

\section{Author contributions}
 M.P. performed the experiment and data analysis, with contributions from J.H. and with assistance from G.P.  The theoretical and experimental conceptualisation was developed by M.G, H.W and G.P.

\section{Competing financial interests}
The authors declare no competing financial interests.


\begin{thebibliography}{99}

\bibitem{background}
Winsberg, E. Computer simulations in science. \textit{Stanford Encyclopedia of Philosophy}, http://plato.stanford.edu/entries/simulations-science/


\bibitem{mgu}
Gu, M.\, Wiesner, K.\, Rieper, E.\, \& Vedral, V.\ Quantum mechanics can reduce the complexity of classical models.
\newblock \textit{Nat. Commun.} \textbf{3,} 762 (2012).


\bibitem{crutch09}
Crutchfield, J.\ P.\, Ellison, C.\ J.\, \& Mahoney, J.\ R.\ Time's barbed arrow: irreversibility, crypticity, and stored information
\newblock \textit{Phys.\ Rev.\ Lett.} \textbf{103,} 094101 (2009).

\bibitem{shor}
Shor, P.\ Algorithms for quantum computation: Discrete logarithms and factoring.
\newblock \textit{Proc. 35th Ann. Symp. on Found. Of Comp. Sci.} 124-134 (1994).

\bibitem{dominic}
Berry, D. W., Childs, A. M., Kothari, R. Hamiltonian simulation with nearly optimal dependence on all parameters. arXiv:1501.01715.

\bibitem{qkd} Gisin, N., Ribordy, G., Tittel, W. and  Zbinden, H. Quantum cryptography. {\it Rev. Mod. Phys.} {\bf 74,} 145-195 (2002).

\bibitem{metrology1} Xiang, G. Y.,  Higgins, B. L., Berry, D. W., Wiseman, H. M. \& Pryde, G. J. Entanglement-enhanced measurement of a completely unknown optical phase. \textit{Nature Photonics} \textbf{5}, 43 (2011).

\bibitem{metrologyrev}
Giovannetti, V.,	 Lloyd, S. \&  Maccone, L. Advances in quantum metrology. \textit{Nature Photon.} \textbf{5}, 222-229 (2011)


\bibitem{Plesch2010}  Plesch, M.\, \& Bu\v{s}ek, V. Efficient compression of quantum information. \textit{Phys. Rev. A} \textbf{81,} 032317 (2010).

\bibitem{Rozema2014} Rozema, L. A.\, Mahler, D. H.\, Hayat, A.\, Turner, P. S.\, \& Steinberg, A. M. Quantum Data Compression of a Qubit Ensemble. \textit{Phys. Rev. Lett.} \textbf{113,} 160504 (2014).

\bibitem{shal01}
Shalizi, C.\ R.\, \& Crutchfield, J.\ P.\ Computational Mechanics: Pattern and Prediction, Structure and Simplicity.
\newblock \textit{Journal of Statistical Physics}\ \textbf{104,} 817 (2001).

\bibitem{hog}
Huberman, B.\, \& Hogg, T.\ Complexity and Adaptation.
\newblock \textit{Physica D} \textbf{22,} 376-384 (1986).




\bibitem{Shal04}
Shalizi, C.\ R.\, Shalizi, K.\ L.\, \& Haslinger, R.\ Quantifying self-organisation with optimal predictors.
\newblock \textit{Phys. Rev. Lett.} \textbf{93,} 118701 (2004).

\bibitem{crutch89}
Crutchfield, J.\ P.\ \& Young, K.\ Inferring statistical complexity.
\newblock \textit{Phys. Rev. Lett.} \textbf{63,} 105 (1989).

\bibitem{chun08}
Chun-Biu Li, T.\ K.\ , \& Yang, H.\ Multiscale complex network of protein conformational fluctuations in single-molecule time series.
\newblock \textit{Proc. Nat. Ac. Sci.} \textbf{105,} 536 (2008).


\bibitem{Schumacher}
Schumacher, B. \& Westmoreland, M. {\em Quantum processes, systems, \& information} (Cambridge University Press, Cambridge, 2010)

\bibitem{karoline}
Wiesner, K.\, Gu, M.\, Rieper, E.\, \& Vedral, V.\ Information-theoretic lower bound on energy cost of stochastic computation.
 \newblock \textit{Proc. of the Roy. Soc. A} \textbf{468}, 2148 (2012)

 \bibitem{andrew}
 Garner, A.\, Thompson J.\, Vedral, V.\, \& Gu, M.\ When is simpler thermodynamically better? \textit{arXiv preprint arXiv:1510.00010} (2015)

 \bibitem{Mahoney2015}
Mahoney, J. R.\, Aghamohammadi, C.\, \& Crutchfield, J. P. Occam's Quantum Strop: Synchronizing and Compressing Classical Cryptic Processes via a Quantum Channel. \textit{arXiv preprint 1508.02760} (2015)


\bibitem{langford}
Langford, N. K., Weinhold, T. J., Prevedel, R., Resch, K. J., Gilchrist, A., O'Brien, J. L., Pryde, G. J. \& White, A. G. Demonstration of a simple entangling optical gate and its use in Bell-state analysis. \textit{Phys. Rev. Lett.} \textbf{95}, 210504 (2005).

\bibitem{white} White, A. G., Gilchrist, A., Pryde, G. J., O'Brien, J. L., Bremner, M. J. \&Langford, N. K. Measuring two-qubit gates. \textit{Journal of the Optical Society of America B} \textbf{24}, 172 (2007).





\end{thebibliography}
\end{document}